\documentstyle [12pt,epsf]{article}

\begin{document}
\baselineskip 7.5 mm

\def\thefootnote{\fnsymbol{footnote}}

\begin{flushright}
\begin{tabular}{l}
UPR-705-T \\
June, 1996 
\end{tabular}
\end{flushright}

\vspace{2mm}

\begin{center}

{\Large \bf 
Velocities of pulsars and neutrino oscillations. 
}
\\ 
\vspace{8mm}

\setcounter{footnote}{0}

Alexander Kusenko\footnote{ email address:
sasha@langacker.hep.upenn.edu; address after October 1, 1996:
Theory Division, CERN, CH-1211 Geneva 23, Switzerland} 
and  
Gino Segr\`{e}\footnote{email address: segre@dept.physics.upenn.edu} 
\\
Department of Physics and Astronomy, University of Pennsylvania \\ 
Philadelphia, PA 19104-6396 \\

\vspace{12mm}

{\bf Abstract}
\end{center}

Neutrino oscillations, biased by the magnetic
field, alter the shape of the neutrinosphere in a cooling protoneutron star
emerging from the supernova collapse.  The resulting anisotropy in the
momentum of outgoing neutrinos can be the origin of the observed proper
motions of pulsars.  
The connection between the pulsars velocities and
neutrino oscillations results in a prediction for the $\tau$ neutrino mass
of $m(\nu_\tau) \sim 100$ eV.

\vfill

\pagestyle{empty}

\pagebreak

\pagestyle{plain}
\pagenumbering{arabic}
\renewcommand{\thefootnote}{\arabic{footnote}}
\setcounter{footnote}{0}

\pagestyle{plain}

It is well known that rotating magnetized neutron stars, pulsars, 
exhibit rapid proper motions~\cite{ll,go} characterized by  space 
velocities that range in the hundreds of kilometers per second.  Born in 
a supernova explosion, a pulsar may receive a substantial ``kick''
velocity due to the asymmetries in the collapse, explosion, and the 
neutrino emission affected by convection~\cite{w,w1}.  Evolution of close
binary systems may also produce rapidly moving pulsars~\cite{b}.
Alternatively, 
it was argued~\cite{hd} that the pulsar may be accelerated during the first
few months after the supernova explosion by its electromagnetic radiation,
the asymmetry resulting from the magnetic dipole moment being inclined to
the rotation axis and offset from the center of the star.  
Most of these mechanisms, however, have difficulties explaining the
magnitudes of pulsar spatial velocities, which can be as great as 
$1000$~km/s and have an average value of $450$~km/s~\cite{ll}.  
(A recent study \cite{w1} shows, however, that a kick velocity of this
magnitude can be achieved in an asymmetric collapse.)

\begin{figure}
\centerline{\epsfbox{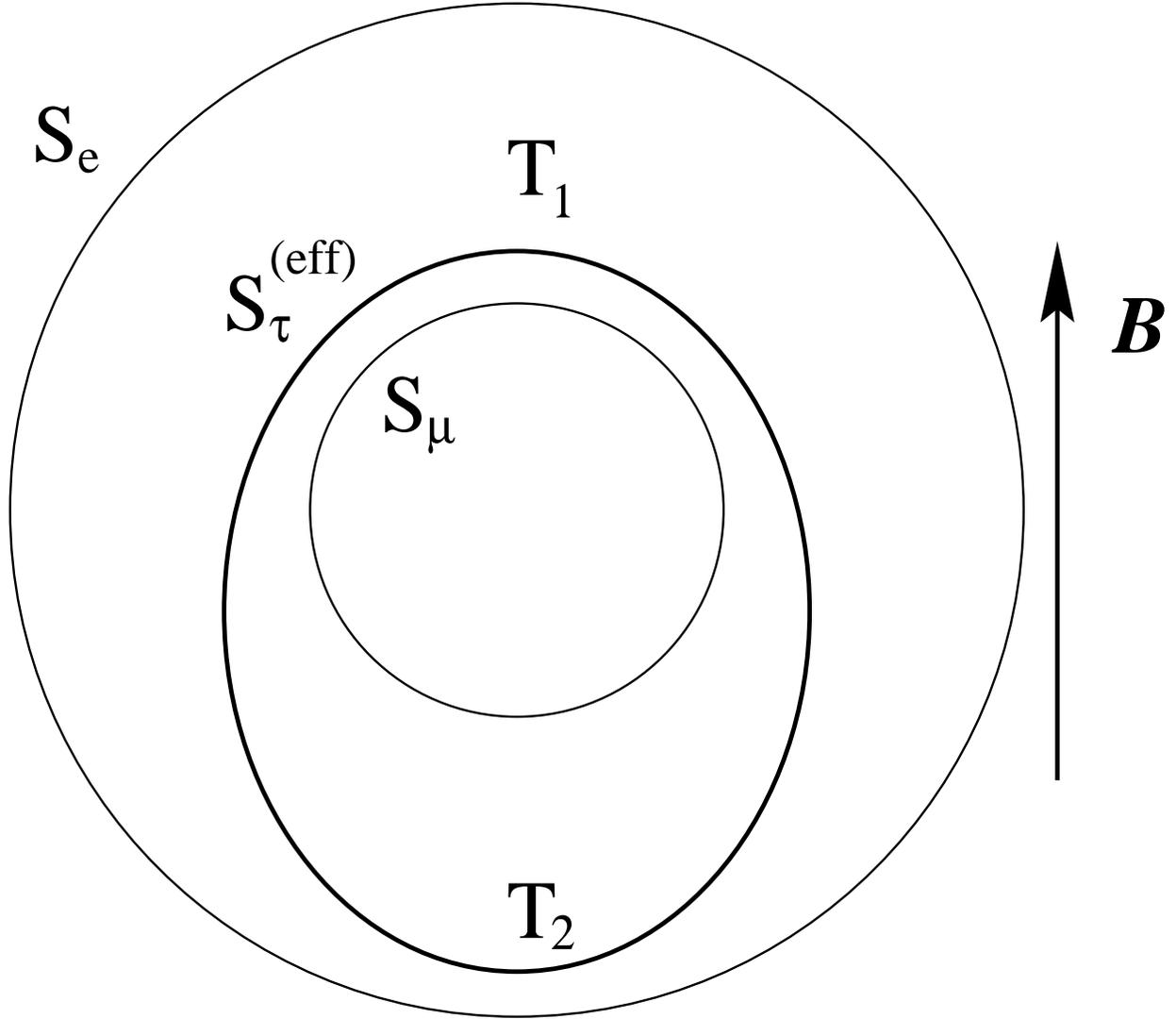}}
\caption{If the resonant oscillation {$\nu_\tau \rightarrow \nu_e$} takes 
place between the electron neutrinosphere, {$S_e$}, and that of 
{$\nu_\mu, \nu_{\tau}, \bar{\nu}_\mu, \bar{\nu}_{\tau}$}  and
{$\bar{\nu}_e$}, {$S_\mu$}, then the {\it effective} 
{$\tau$}-neutrinosphere coincides with the surface of resonance,
{$S^{({\rm eff})}_\tau$}.  
The latter is affected by the magnetic field.  Therefore, the 
{$\tau$}-neutrinos emitted in different directions come from the regions of 
different temperatures.  The resulting anisotropy in the momentum of the
outgoing neutrinos can be the origin of the pulsar ``kick'' velocity, whose
magnitude depends on the magnetic field.}
\label{fig1}
\end{figure}

In this letter we suggest an explanation for the birth velocities of
the neutron stars based on asymmetric emission of neutrinos due to neutrino
oscillations.  We also show that
the distribution of pulsar velocities can yield information about neutrino
masses. 

The basic idea is the following.  Neutrinos emitted during the cooling of
a protoneutron star have  total momentum, roughly, $100$ times
the momentum of the proper motion of the pulsar.  A $1$\% 
anisotropy in the neutrino distribution would result in a ``kick''
velocity consistent with observation.   In the dense neutron star an
electron neutrino, $\nu_e$, has a shorter mean free path than $\nu_\mu$,
$\nu_\tau$, or any of the antineutrinos.  If one of the latter, {\it
e.\,g.}, $\nu_\tau$, undergoes a resonant oscillation into $\nu_e$, above 
the $\tau$-neutrinosphere but below the $e$-neutrinosphere, it will be
absorbed by the medium.  Therefore, the {\it effective}
$\tau$-neutrinosphere in this case is determined by the point of resonance
(Fig \ref{fig1}).    

It is known that the electromagnetic properties of a neutrino propagating
in medium, in a longitudinal magnetic field $\vec{B}$, are different from 
those in a 
vacuum \cite{nu_in_medium}.  The effective neutrino self-energy has a
contribution proportional to $(\vec{B} \cdot \vec{k})$, where $\vec{k}$ is 
the neutrino momentum.  The position of the resonance of the
$\nu_\tau  \rightarrow \nu_e$ oscillations is affected by the magnetic
field and depends on the relative orientation of $\vec{B}$ and $\vec{k}$ 
\cite{nu_osc_B}.  Therefore the effective
$\tau$-neutrinosphere (or, ``neutrinosurface'', to be more precise) is not
concentric with the electron neutrinosphere (and, in fact, is not a
sphere).  The $\tau$ neutrinos that escape in the direction of magnetic
field have, therefore,  a different temperature from those emitted in the
opposite direction.   They carry away  different momenta, thereby creating
an asymmetry in good quantitative agreement  with the data. 
We will show that this mechanism may be the origin of
the proper motions of pulsars.

In a recent analysis of a sample of 99 pulsars, Lyne and Lorimer \cite{ll} 
concluded that the space velocity of pulsars at birth has a mean value of 
$450 \pm 90 $ km s$^{-1}$.  Typical pulsars have masses in the range from
$1.0 \ M_{\odot} $ to $1.5 \ M_{\odot} $, where $M_{\odot}$ is the
solar mass \cite{pulsar_review}.  The momenta associated with the
proper motion of pulsars are therefore of order 
$ k_p \approx (1.2\pm 0.4)\times 10^{41} \: {\rm g \, cm/s } $. 
The energy carried off by neutrinos in a supernova explosion is estimated 
to be $\sim 3\times 10^{53}$ erg \cite{snu_review}.  Since $p_\nu \approx
E_{\nu}$ (we set $\hbar=c=1$), this corresponds to the sum of the
magnitudes of the neutrino momenta $\approx 10^{43}$ g cm/s.  Comparing this
with $k_p$, we conclude that  a few per cent asymmetry in the
distribution of the outgoing neutrinos is sufficient to give the pulsar a
``kick'' velocity  $\approx 450$ km s$^{-1}$.

The average energy of the neutrinos emitted during the cooling of the 
protoneutron star depends on the temperature of the neutrinosphere
\cite{snu_review}, defined as the surface where the optical depth,
$\int_R^\infty dr/\lambda_\nu$, becomes of order $1$.  Neutrinos of
different types have, in general, different neutrinospheres because of the
difference in the mean free path, $\lambda_\nu \propto 1/\sigma_\nu$.
Roughly speaking,  
one can consider the inside of the neutrinosphere to be opaque for
neutrinos, while the outside is transparent.  Electron neutrinos 
have both charged and neutral current interactions;  they 
scatter off electrons, positrons and nucleons, as well as being absorbed 
\cite{sa} in the reaction $\nu_e n \rightarrow e^{-}p$. On the other hand, 
$\mu$ and $\tau$ neutrinos and antineutrinos can only scatter
elastically, via neutral currents, at  MeV temperatures.  
(The electron antineutrino, $\bar{\nu_e}$ can scatter and be absorbed 
in the process $\bar{\nu_e} p \rightarrow e^{+}n$, so its opacity starts
out close to that of $\nu_e$ and then decreases to, roughly, that of
$\nu_\mu$ and $\nu_\tau$ as the density of protons decreases.) 

This difference in scattering cross-sections gives rise to roughly an
order of magnitude difference \cite{sa,bl} in the mean free paths of the
$\nu_e$ and the $\nu_\tau$ in the vicinity of the electron neutrinosphere,
where the density is $10^{11}-10^{12}$ g/cm$^3$.  Consequently, the
$\tau$-neutrinosphere lies deeper inside the protoneutron star than the
electron neutrinosphere.  As a result, the $\tau$ neutrinos escape from a
deeper layer of the star, where the temperature is higher, and their mean
energy exceeds that of $\nu_e$'s by about 50\%. 

We now consider neutrino oscillations.  For definiteness, we discuss 
two flavors, $\nu_e$ and $\nu_\tau$, with $\Delta m^2 \equiv
m^2(\nu_\tau)-m^2(\nu_e)  
\approx m^2(\nu_\tau) \gg m(\nu_e) \approx 0$ and small mixing.  We will
also assume that  $\nu_e \rightarrow \nu_\mu$ oscillations  
take place in the Sun, while the transition $\nu_\tau \rightarrow \nu_e$
occurs at a much higher matter density.  We leave a more careful analysis
of the neutrino mass matrix consistent with all experimental data for
future publication \cite{future}.

As was shown in Ref. \cite{nu_osc_B}, the neutrinos of energy $E\approx 
k = |\vec{k}|$ propagating in the 
degenerate electron gas of the charge density $N_e=n_{e^{-}}-n_{e^{+}}$, in
magnetic field $\vec{B}$, undergo a resonant oscillation\footnote{
We emphasize the difference between these chirality-preserving oscillations
and the spin and flavor precession of neutrinos in magnetic field studied,
{\it e.\,g.}, in Ref. \cite{fs}.  The last term in equation (\ref{res})
arises due to the effective interaction of neutrinos with the charged
particles in the background and does not depend on the size of the neutrino
magnetic dipole moment.} if  

\begin{equation}
\frac{\Delta m^2}{2 k} \: cos \, 2\theta = \sqrt{2} \, G_{_F} \, N_e +
\frac{e G_{_F}}{\sqrt{2}} \left ( \frac{3 N_e}{\pi^4} \right )^{1/3} \ 
\frac{\vec{k} \cdot \vec{B}}{k},
\label{res}
\end{equation}
where $\theta$ is the neutrino mixing angle in vacuum.

If the resonant conversion of $\nu_\tau$ into $\nu_e$ occurs at some point
between the $\tau$ neutrinosphere and the electron neutrinosphere, the 
$\tau $ neutrinos produced inside the ``resonance-sphere'' ({\it i.\,e.} 
closer to the center than the point of the resonance), will turn into
$\nu_e$'s and will be absorbed (thermalized) by the medium.  
The $\tau$-neutrinos produced by the $\nu_e \rightarrow \nu_\tau$
oscillations will escape with the energy determined by the temperature at
the resonance point.  Of course,  $\tau$-neutrinos produced outside the
resonance-sphere will also escape, but there is no effective mechanism for
copious production of $\tau$-neutrinos outside neutrinosphere, where the
matter density is small.  The surface of resonance points for the $\nu_\tau
\rightarrow \nu_e$ oscillations  becomes the effective $\tau$
neutrinosphere.  Since the last term in (\ref{res}) depends on the
relative orientation of $\vec{k}$ and $\vec{B}$, the resonant
oscillations occur at different densities for the neutrinos going in 
different directions. The corresponding difference in
temperatures will result in the asymmetry that gives the neutron star a
``kick'' in the direction of the magnetic field.

In the absence of the magnetic field, the resonance will occur at a 
distance $r_0$ from the center if 

\begin{equation}
\Delta m^2 \: cos \, 2 \theta = 2\sqrt{2} \, G_{_F} \, N_e(r_0) \, k.
\end{equation}

If $r_0$ is to lie between the two neutrinospheres, where the density is 
$\approx 10^{11}$ g cm$^{-3}$ and the electron to baryon ratio 
$Y_e \approx 0.1$, then, for a small mixing angle, 
$\Delta m^2$ has to be of order $10^4$ eV$^2$, which corresponds to
a $\tau$ neutrino mass of $100$ eV, consistent with
the current experimental limits \cite{pdg}.  This is close
to the limit on stable neutrino masses imposed by cosmological constraints,
$m_\nu <92 \ {\rm eV} \ (\Omega h_0^2)$ \cite{dm}, although the mixed dark
matter models favor a lower mass range \cite{dm_favored} for stable
neutrinos.   Larger values of mass have been considered for unstable
$\nu_\tau$'s \cite{dgt}.  

The oscillations are adiabatic as long as the density $N_e(r_0)$ and the
magnetic field can be considered constant over the oscillation length,
$l_{osc}$, 

\begin{equation}
l_{osc} \approx \left (\frac{1}{2\pi} \ 
\frac{\Delta m^2}{2 k} \ sin \, 2 \theta
\right )^{-1} \approx \frac{1 \: {\rm cm}}{sin \, 2 \theta }.
\end{equation}
For a wide range of mixing angles, $l_{osc}$ is much smaller than the scale
on which the density changes are noticeable, so the oscillations can, in
fact, be treated as adiabatic. 

In the presence of the magnetic field, the condition (\ref{res}) is
satisfied at different distances from the center, depending on the value of
the  $(\vec{k} \cdot \vec{B})$ term in (\ref{res}). The surface of the
resonance is, therefore, 

\begin{equation}
r(\phi) = r_0 + \delta \: cos \, \phi, 
\end{equation}
where $cos \, \phi= (\vec{k} \cdot \vec{B})/k$ and $\delta$ is determined
by the equation: 

\begin{equation}
2 \frac{d N_e(r)}{dr} \delta \approx 
e \left ( \frac{3 N_e}{\pi^4} \right )^{1/3} B. 
\end{equation} 
This yields

\begin{equation}
\delta = \left ( \frac{3 N_e}{\pi^4} \right )^{1/3} \:
\frac{e}{2} \: B \left / \frac{dN_e(r)}{dr} \right. =
\frac{e \mu_e}{2 \pi^2} \: B \left / \frac{dN_e(r)}{dr} \right. ,
\label{delta}
\end{equation}
where $\mu_e \approx (3 \pi^2 N_e)^{1/3} $ is the chemical potential of the
degenerate (relativistic)  electron gas.

We now estimate the size of the ``kick'' velocity from the effect just
described.  As was established in the beginning, a few per cent  
asymmetry in the momentum distribution of neutrinos is necessary to explain
the observed pulsar velocities.  Approximately $1/6$ of the total energy is
carried off by each of the neutrino species \cite{snu_review}.  Integration
over the angles in $(\vec{k} \cdot \vec{B})$ gives a factor $1/2$ if the
magnetic field is uniform.  (The latter is, of course, a simplification.)
The asymmetry in the third component of momentum is, therefore,  
\begin{equation}
\frac{\Delta k}{k} = \frac{1}{6} \ \frac{1}{2} \ 
\frac{T^4(r_0-\delta )-T^4(r_0+\delta)}{T^4(r_0)} \approx
\frac{2}{3} \ \frac{1}{T} \ \frac{dT}{dr} \ \delta,
\end{equation}
where we have assumed a black-body radiation luminosity $\propto T^4$ for
the effective neutrinosphere.  Using the expression (\ref{delta}) for
$\delta$, we obtain

\begin{equation}
\frac{\Delta k}{k} = \frac{e}{3 \pi^2} \: \left ( \eta  \frac{dT}{dN_e}
\right) B, 
\label{dk}
\end{equation}
where $\eta \equiv \mu_e/T$ is the degeneracy parameter of the electrons.
Here we used the identity $(dT/dr)/(dN_e/dr) \equiv dT/dN_e$. 
To calculate the derivative in (\ref{dk}), we use the relation between the
density and the temperature for the relativistic Fermi gas: 

\begin{equation}
N_e=2 \int \frac{d^3p}{(2\pi)^3} 
\frac{1}{e^{(p-\mu)/T}+1}.
\label{fermi}
\end{equation}
Differentiating the right-hand side with respect
to $T$, we obtain
$ dN_e/dT  
\approx (2/3) \, T^2 \, \eta$, 
for large $\eta$.  (This asymptotic expression is accurate to three
significant digits for $\eta>5$.)
We observe that for a degenerate relativistic electron gas ($\eta \gg 1$)
the product $ \eta (dT/dN_e) \approx 1.50/T^2$, independent of $\eta$.
Finally, the ratio in (\ref{dk}) is 

\begin{equation}
\frac{\Delta k}{k}= 0.015 \ \frac{B}{T^2} = 0.01 \left ( 
\frac{3 \ {\rm MeV}}{T} \right )^2
\left ( \frac{B}{3 \times 10^{14} G} \right ).
\label{final}
\end{equation}

During the cooling stage of the protoneutron star, the $\tau$-neutrinos
come out with the average energy $\approx 10$ MeV \cite{snu_review}, which
corresponds to the temperature of $\approx3$ MeV.  We see that the observed
pulsar velocities, which require $\Delta k/k$ to be of order $0.01$,
can be explained by the values of $B \sim 3\times 10^{14}$ G. 

The magnetic fields at the surface of the neutron stars are estimated to be
of order $10^{12}-10^{13}$G \cite{pulsar_review}.   However, a magnetic 
field inside the pulsar may be as high as $10^{16}$G 
\cite{pulsar_review,r}.  The existence of such a strong magnetic field is
suggested by the dynamics of formation of the neutron stars, by the 
stability of the poloidal magnetic field outside the pulsar, as well as by 
the fact that the only star whose surface field is well studied, the Sun,
has magnetic field below the surface which is $10^3$ larger than that
outside. (This strong magnetic field causes the sun spots when it
penetrates the surface.)  

Magnetic fields of order $10^{16}$G inside the
neutron star can themselves lead to asymmetric neutrino flux by 
the creation of neutron star analogue of sun spots.  This phenomenon is
disscussed in the context of strongly magnetized neutron starts 
by R.~C.~Duncan and C.~Thompson \cite{dt}.  In such strong magnetic 
fields, other weak interactions effects also become important \cite{doro}.  
We emphasize that the magnetic fields relevant for these effects 
are an order of magnitude higher than those considered in the present 
work. 

It is clear from equation (\ref{final}) that the deformations of the
neutrinosphere due to neutrino oscillations biased by the magnetic field 
can result in the asymmetry of the neutrino flux necessary to give the
pulsar a ``kick'' velocity consistent with the data.

In our calculations, we have assumed that the 
temperature distribution is not
affected significantly by the position of the resonance.  This assumption
is well-justified because only the neutrinosphere of one out of six 
(anti-)neutrino species depends on the resonance, while the temperature 
profile is determined by the emission of all six.  However, the increase of
the heat drain in the vicinity of the effective  $\tau$-neutrinosphere 
makes the temperature gradient larger. This effect further increases the
ratio (\ref{final}).  We have also neglected the deviations from thermal
equilibrium in a cooling protoneutron star.  A detailed analysis of the
heat transport equations, though clearly important, is beyond the scope of
this letter.  

According to equation (\ref{final}), the birth velocities of neutron stars 
are proportional to the magnetic field at the effective neutrinosphere.
However, the direction of the proper motion of a pulsar need not be
correlated with its angular momentum, because of the inclination of the
magnetic field to the rotation axis and its possible offset from the center. 
This is also in good agreement with the observations which find no 
correlation between the space velocities and the linear polarization 
of radio emission \cite{al}.  In a separate paper \cite{ks2}, we discuss
the implications of our mechanism for the correlation between the magnetic
fields and the magnitudes of pulsar velocities.  

In conclusion, we have described a new explanation for the 
birth velocities of pulsars, which is in good agreement with
the observational data. It presupposes, as we have stated, a $\tau$
neutrino mass of about 
$100$  eV for small mixing.  (A more complex realization of our mechanism
may allow for smaller $m(\nu_\tau)$ \cite{future}.)   
The surprising connection between the proper motions of pulsars and
neutrino oscillations provides a new astrophysical ``laboratory'' for
neutrino physics.  Measurements of the space velocities of pulsars can
yield information about the neutrino masses in the  range which is
currently inaccessible to either solar neutrino, or collider experiments.   

We would like to thank A.~Dolgov, V.~Kaspi, D.~Lai,  P.~Langacker,  
M.~Ruderman, 
R.~Sawyer, D.~Schramm, D.~Seckel, R.~Shrock, A.~Smirnov and L.~Wolfenstein
for helpful discussions.  This work was supported by the 
U.~S.~Department of Energy Contract No. DE-AC02-76-ERO-3071.

\end{document}